
  \hsize=17 true cm
  \parindent=0 true cm


  \def\sa{\vskip 0.30 true cm}
  \def\sb{\vskip 0.60 true cm}
  \def\sc{\vskip 0.15 true cm}

  \def\aa{\vskip 0.30 true cm}
                          
  \baselineskip = 0.60 true cm

\rightline{\bf {LYCEN 9009}}

\rightline{(March 1990)}

\sa
\sb
\sa
\sb

\centerline{\bf ON AN ALTERNATIVE PARAMETRIZATION FOR} 

\sc

\centerline{\bf THE THEORY OF COMPLEX SPECTRA}

\sa
\sb

\centerline{{\bf M. KIBLER}\footnote*{Permanent address : 
Institut de Physique Nucl\'eaire de 
Lyon, IN2P3-CNRS et Universit\'e Claude Bernard, F-69622 
Villeurbanne Cedex, France.} and {\bf J. KATRIEL}}

\sa

\centerline{Department of Chemistry,} 

\centerline{Technion \ - \ Israel Institute of Technology,}

\centerline{Haifa 32000, Israel} 

\sa
\sb
\sb
\sb

\parindent = 1 true cm
\baselineskip = 0.72 true cm

\centerline{ABSTRACT}

\sa

The purpose of this letter is threefold : 
(i) to derive, in the framework of a new parametrization, some 
compact formulas of energy averages for the electrostatic 
interaction within an $n {\ell}^N$ configuration, 
(ii) to describe a new generating function for obtaining 
the number of states with a given spin angular momentum in an 
$n {\ell}^N$ configuration, 
and (iii) to report some apparently new sum rules (actually a 
by-product of (i)) for $SU(2) \supset U(1)$ coupling coefficients. 

\sa
\sb
\sb
\sb
\sb
\sb
\sb

Published in Physics Letters A {\bf 147}, 417-422 (1990).

\vfill\eject

\parindent = 1 true cm
   \baselineskip = 0.88 true cm

\noindent{\bf 1. Introduction}

\sa

In the theory of complex spectra [1-3], one- and two-body Hamiltonians 
invariant under the group $SO(3)$ and symmetric in the spin and orbital parts 
can be written in the form [4-5] 
$$
V = \sum_{i \not = j} \sum_{{\rm all} \, k} 
\; D[(k_1k_2)k_S(k_3k_4)k_L] \; 
\{
\{u^{(k_1)}(i) \otimes u^{(k_2)}(j)\}^{(k_S)} \otimes 
\{u^{(k_3)}(i) \otimes u^{(k_4)}(j)\}^{(k_L)}
\}^{(0)}_0.
\eqno (1)
$$
In eq.~(1), $u^{(k)}$ stands for a Racah unit tensor of rank $k$. 
Further, the $SO(3)$-invariant operator $\{ \ \ \}^{(0)}_0$ results 
from the coupling of the
tensor product $\{ u^{(k_1)}(i) \otimes u^{(k_2)}(j) \}^{(k_S)}$ 
acting on the spin part with the tensor product 
               $\{ u^{(k_3)}(i) \otimes u^{(k_4)}(j) \}^{(k_L)}$
acting on the orbital part. Finally, the $D[ \ \ ]$ parameters in eq.~(1) are
radial parameters (depending on the radial wave\-functions involved, e.g., 
$R_{n \ell}(r)$, $R_{n'\ell'}(r)$, etc.) which are generally taken as
phenomenological parameters.

\sa

The operator (1) can be considered as a particular case of the $G$-invariant
Hamiltonian introduced in [5] for describing optical and magnetic
properties of a partly-filled shell ion 
in a crystalline environment with symmetry
$G$. Equation (1) corresponds to $G \equiv SO(3)$ : to obtain (1) from 
ref.~[5], it 
is enough to put $k = 0$ and to replace $a_0$ or $a_0 \Gamma_0 \gamma_0$ 
by $q = 0$.

\sa

We shall be concerned in this work with the (spin-independent) 
Coulomb interaction which is
obtained from (1) by taking $k_1 = k_2 = 0$ and $k_3 = k_4 = k$. In this case,
the parameters $D[(00)0(kk)0]$ are proportional to the Slater parameters
$F^{(k)}$. We shall restrict ourselves to the action of the Coulomb 
interaction $V(k_1 = k_2 = 0)$ within an $n\ell^N$ configuration but, for
the purpose of forthcoming generalizations, we shall consider Slater 
parameters of the type 
$F^{(k)}(\ell, \ell') = R^{(k)}(\ell, \ell'; \ell, \ell')$. 
The $F^{(k)}(\ell, \ell')$
parametrization corresponds to a multipolar expansion of the electrostatic 
interaction $V(k_1 = k_2 = 0)$ and turns out to be especially adapted
to the chain $SO(3) \supset SO(2)$. 

\sa

There are several other parametrizations besides the $F^{(k)}(\ell, \ell')$
parametrization. (For instance, 
the $E^k$ parametrization [3] is well-known 
for $nf^N$ configurations.) An alternative parametrization, referred 
to here as
the the ${\cal E}^\lambda (\ell, \ell')$ parametrization, was introduced in 
ref.~[6]. This parametrization was obtained by transposing to atomic and
nuclear spectroscopy a parametrization, namely, the Angular Overlap Model 
parametrization [7,8], used in the spectroscopy of partly-filled shell ions
in crystalline environments. The ${\cal E}^\lambda (\ell, \ell')$
parametrization can be defined by the passage formula
$$
F^{(k)}(\ell, \ell') = 
{{2 k + 1} \over {2\ell +1}} \, 
{\pmatrix{
\ell & k & \ell'\cr
   0 & 0 &     0\cr
}}^{-1}
\sum _\lambda \, (-1)^{\lambda}
\pmatrix{
\ell     & k & \ell'\cr
-\lambda & 0 & \lambda\cr
} 
{\cal E}^\lambda(\ell, \ell')
\eqno (2)
$$
or the reverse formula
$$
{\cal E}^\lambda(\ell, \ell') =
(-1)^\lambda \sqrt {(2\ell + 1) (2\ell' +1)}\;
\sum _k 
\pmatrix{
\ell & k & \ell'\cr
0 & 0 & 0\cr
}
\pmatrix{
\ell & k & \ell'\cr
-\lambda & 0 & \lambda\cr
}
F^{(k)}(\ell, \ell'),
\eqno (3)
$$
which generalizes eq.~(2) of ref.~[6].

\sa

It is one of the aims of this letter to show (in section 3) that various energy
averages for $n\ell^N$ configurations assume a particularly simple form when
expressed in the ${\cal E}^\lambda (\ell, \ell')$ parametrization. To 
obtain 
energy averages, one needs to know the number of states, 
having well-defined qualifications, in the $n \ell^N$ 
configuration and a new way of denumbering
states, originally introduced in ref.~[9], is further developed in section 2.
Two new sum rules for $3-jm$ symbols, which are at the root of the derivation
of two energy averages in section~3, are relegated to an appendix. Finally, 
some concluding remarks are given in section~4.

\sa
\sa

\noindent {\bf 2. Denumbering states}

\aa

Let us consider a system of $N$ fermions in a shell $n\ell$. As is well-known,
the resulting configuration $n\ell^N$ has $4\ell +2\choose N$ totally
anti-symmetric state vectors. Among these $4\ell + 2\choose N$ states, let
$H_{\ell}(N,S)$ be the number of states having a given total spin $S$. We
devote this section to the calculation of $H_{\ell}(N,S)$.

\aa

From ref.~[9], we know that the function
$$
F(x,y,z) = {\prod^{1/2}_{m_s=-1/2}} \ 
{\prod^{\ell}_{m_{\ell} = - \ell}} \ 
(1 + z \> y^{m_s} x^{m_{\ell}})
\eqno (4)
$$ 
is the generating function for the number $F_{\ell}(N,M_S,M_L)$ of states in
$n\ell^N$ with $z$-compo\-nents of the total spin and orbital 
angular momenta equal to
$M_S$ and $M_L$, respectively. The number $F_{\ell}(N,M_S,M_L)$ is obtained by
expanding $F(x,y,z)$ as
$$
F(x,y,z) = \sum_{N \, , \, M_S, \, M_L} \ 
F_{\ell}(N,M_S,M_L) \; z^N \; y^{M_S} \; x^{M_L}. 
\eqno (5)
$$
By setting $x = 1$, we obtain a generating function for 
$$
G_{\ell}(N,M_S) = \sum_{M_L} \; F_{\ell}(N,M_S,M_L)
\eqno (6)
$$
which is the number of states with a definite value of $M_S$. The latter
generating function can be written in two equivalent forms, namely :
$$
(1 + z \, y^{-1/2})^{2\ell +1} \ (1 + z \, y^{1/2})^{2\ell +1} 
\qquad {\rm or } \qquad 
\left [1 + z \, (y^{-1/2} + y^{1/2}) + z^2 \right ] ^{2\ell + 1}.
\eqno (7)
$$
This leads to the following expression
$$
G_{\ell}(N,M_S) =
\pmatrix{
2\ell&+&1\cr 
{N \over 2}&-&M_S\cr 
}
\pmatrix{
2\ell&+&1\cr
{N \over 2}&+&M_S\cr
},
\eqno (8)
$$
or alternatively
$$
G_{\ell}(N,M_S) = (2\ell + 1)! \, 
\sum_{i = 0}^{[N/2]} \, 
{1 \over {i! \; (2\ell + 1 - N + i)! \; ({N\over 2} - i - M_S)! \; 
({N\over 2} - i + M_S)!}}. 
\eqno (9)
$$
In particular, the compact form (8) is very simple to handle. Finally, the
number $H_{\ell}(N,S)$ of states of the configuration $n\ell^N$ with total 
spin $S$ is simply obtained by combining
$$
H_{\ell}(N,S) = G_{\ell}(N,S) - G_{\ell}(N,S + 1) 
\quad {\rm for} \quad S < {N \over 2}, 
$$
$$
H_{\ell}(N,{N \over 2}) = G_{\ell} (N,{N \over 2}) 
\quad {\rm for} \quad S = {N \over 2}
\eqno (10)
$$
with eq.~(8).

\aa

As an illustration, we consider the configuration $nf^6$. From 
eq.~(8), the
numbers $G_3(6,M_S)$ are found to be $7$, $147$, $735$ and $1225$ 
for $M_S = 3$, 2, 1 and $0$, respectively. Furthermore from 
eq.~(10), the numbers $H_3(6,S)$ are 
easily seen to be $7$, $140$, $588$ and $490$ for $S = 3$, 2, 1 and $0$,
respectively. As a check, we verify that the total number of states 
is $3003 = {14 \choose 6}$.

\aa
\aa

\noindent {\bf 3. Averages for $n\ell^N$ configurations}

\aa

{\it 3.1. Average interaction energy}

\aa

The average interaction energy $E_{av}(\ell ,\ell)$ for an arbitrary 
$n\ell^N$ configuration in spherical symmetry was derived by 
Shortley [2] and 
further discussed by Slater in his book [1] 
(see also the nice book by Condon and Odaba\c s\i \  [2]). The 
expression 
for $E_{av}(\ell , \ell)$ is known in terms of the Slater parameters $F^k(\ell
, \ell)$. The formula for $E_{av}(\ell ,\ell)$ in the $F^k(\ell , \ell)$
parametrization [1,2] does not exhibit any remarkable  peculiarity. It is a 
simple matter of calculation (by means of Wigner-Racah calculus) 
to convert the expression
for $E_{av}(\ell , \ell)$ in the ${\cal E}^\lambda(\ell ,\ell)$ parametrization.
This yields
$$
E_{av}(\ell ,\ell) = {1\over {4\ell +1}} \; 
{{N(N - 1)} \over 2} \; 
\left( {\cal E}^\sigma + 4 \; {\sum_{\lambda = 1}^\ell} \; 
{\cal E} ^\lambda \right).
\eqno (11)
$$
It is to be noted that, from a fitting procedure viewpoint, eq.~(11) 
involves two parameters (${\cal E}^\sigma \equiv {\cal E}^0$ 
and ${\sum^\ell_{\lambda =1}}\;
{\cal E}^\lambda)$. This inclines us to introduce the linear combinations
$$
{\cal S} = {1\over \ell} \; {\sum^\ell_{\lambda = 1}} \;
{\cal E}^\lambda , \qquad
{\cal D} = {1\over {\ell +1}} \; ({\cal E}^\sigma - {\cal S}).
\eqno (12)
$$
Thus, eq.~(11) can be rewritten as 
$$
E_{av}(\ell ,\ell) =
{{N(N-1)}\over 2} \; 
\left( {\cal S} + {{\ell + 1} \over {4 \ell + 1}} \; {\cal D} \right),
\eqno (13)
$$
in terms of the non-independent parameters ${\cal S}$ and ${\cal D}$.

\aa
 
In the special case where all the parameters ${\cal E}^\lambda$ are taken to be
equal, say to a test value ${\cal E}$ (i.e., 
${\cal S} = {\cal E}$ and 
${\cal D} = 0)$, equations $(11)$ and (13) lead to 
$$
E_{av} (\ell , \ell) = {{N(N - 1)} \over 2} \; {\cal E}.
\eqno (14)
$$
Such a kind of result is especially important for the purpose of checking
electrostatic interaction matrices. Indeed, in the case where ${\cal E}^\lambda
= {\cal E}$ for any $\lambda$, it can be shown that all energy levels have the
same value, viz., $[N(N - 1)/2] {\cal E}$ ; then, 
the spectrum of $V(k_1 = k_2 = 0)$ is maximally 
degenerate and $(14)$ is a simple consequence of this maximal degeneracy.

\aa

{\it 3.2. Other average energies}

\aa

We now turn our attention to the average energy $^{2S + 1}E(\ell ,\ell)$ over
all the states of the configuration $n\ell^N$ corresponding to a given total
spin $S$. From extensive calculations 
of the electrostatic energy levels for the electronic 
configurations $np^N$, $nd^N$ and $nf^N$, we empirically discovered that 
$E_{av}(\ell ,\ell)$ is given by
$$
^{2S + 1}E_{av}(\ell , \ell) =
{{N (N - 1)} \over 2} \;
{\cal S} + {1 \over 2} 
\left [ 
{N \over 2}({N \over 2} + 1) - S(S + 1)
\right ]
{\cal D}.
\eqno (15)
$$
(All matrix elements of the Coulomb interaction $V(k_1 = k_2 = 0)$ for the 
configurations $np^N$, $nd^N$ and $nf^N$ listed in ref.~[10], in the $E^k$ 
(Racah) parametrization for $\ell = f$ and in the 
$F^{(k)}(\ell, \ell')$ (Slater) parametrization 
for $\ell = d$ and $p$, have been transcribed in the 
${\cal E}^\lambda (\ell, \ell')$ parametrization with the help of the 
algebraic and symbolic programming system REDUCE. The complete electrostatic 
energy matrices for the electronic 
configurations $np^N$, $nd^N$ and $nf^N$ in the 
${\cal E}^\lambda (\ell, \ell')$ parametrization are presently under 
preparation for distribution to the interested readers.) 
The case $S = {N \over 2}$ is of special interest ; 
in this case, which corresponds to the highest multiplicity term, eq.~(15) 
simply reduces to 
$$
^{N + 1}E_{av} (\ell , \ell) =
{{N(N - 1)} \over 2} \; {\cal S},
\eqno (16)
$$
so that $^{N + 1}E_{av}(\ell , \ell)$ does not depend on the parameter 
${\cal E}^\sigma$.

\aa

Here again, we note that in the special case where ${\cal E}^\lambda = {\cal
E}$ for any $\lambda$, we obtain from (15)
$$
^{2S + 1}E_{av} (\ell , \ell) =
{{N(N-1)}\over 2}\; {\cal E},
\eqno (17)
$$
a result to be compared with eq.~$(14)$.

\aa

The proof of eq.~(15) can be achieved by constructing all determinental 
wave\-functions with a given value $M_S$ of the $z$-component of the total spin
angular momentum and by calculating the sum of the energies of all these
determinental wave\-functions. The complete proof shall be reported elsewhere.
The proof of (15) for $N > 2$ can be also obtained, in principle, as an
extension of the one for $N = 2$ and we now  derive eq.~(15) for an 
$n \ell^2$ configuration.

\aa

We start from the relation $(3)$ of ref.~[6] 
giving the energy of the term
$^{2S + 1}L$ of the configuration $n\ell^2$. Such a relation can be rewritten
as
$$
^{2S + 1}L = (2 \ell + 1)\; 
\sum^\ell_{\lambda = - \ell}\;
\pmatrix{
\ell &\ell & L \cr
0 & \lambda & - \lambda \cr
}^2 \;
{\cal E}^\lambda (\ell , \ell).
\eqno (18)
$$
Therefore, for the configuration under consideration, we have
$$
^{2S + 1}E_{av}(\ell , \ell) = 
{1\over {\sum_{L_\pi}}\; (2L_\pi + 1)}\;
\sum_{L_\pi} \; (2L_\pi + 1)\;
^{2S + 1}L_\pi,
\eqno (19)
$$
where the sums over $L_\pi$ are to be performed on even values of 
$L$, $L_\pi = 0 (2) (2\ell)$, or on odd values of 
$L$, $L_\pi = 1 (2) (2\ell - 1)$, 
according to whether $S$ is 0 (singlet states) or 1 (triplet states). (As 
usual, the notation $i = a(b)c$ means that $i$ takes the values 
$a$, $a + b$, $a + 2b$, $\dots$, $a + [{{c-a} \over b}]b$.) By 
combining eqs.~(18) and $(19)$, we get
$$
^{2S + 1}E_{av}(\ell , \ell) =
{{2\ell + 1} \over {\sum_{L_\pi}} \; (2L_\pi + 1)} \; 
\sum^\ell_{\lambda = - \ell}\;
{\cal E}^\lambda (\ell, \ell) \;
\sum_{L_\pi M} \;
(2 L_\pi + 1) \;
\pmatrix{
\ell &\ell & L_\pi \cr
0 & \lambda & M \cr
}^2.
\eqno (20)
$$
Although there is a well-known formula for expressing the last sum in $(20)$
when $\sum_{L_\pi}$ is replaced by $\sum_L$ with 
$L = 0 (1) (2\ell)$, to the
best of our knowledge there is no formula in the literature for calculating the
last sum in $(20)$ for the two distinct cases where 
$L_\pi = 0 (2) (2 \ell)$ and 
$L_\pi = 1 (2) (2 \ell - 1)$. By using the formula (30) derived in the
appendix, the sum $\sum_{L_\pi M}$ in $(20)$ can be calculated and we
finally arrive at
$$
^1E_{av}(\ell , \ell) =
{1\over {\ell + 1}} \;
\left [
{\cal E}^\sigma (\ell , \ell) +
\sum^\ell_{\lambda = 1} \;
{\cal E}^\lambda (\ell , \ell)
\right ]
\eqno (21)
$$
for the singlet states and
$$
^3E_{av} (\ell , \ell) =
{1\over \ell} \;
\sum^\ell_{\lambda = 1} \;
{\cal E}^\lambda (\ell , \ell)
\eqno (22)
$$
for the triplet states. It is immediate to check that eqs.~(21) and $(22)$
are particular cases of the general formula $(15)$ 
corresponding to $(N = 2, \; S = 0)$ and $(N = 2, \; S = 1)$, respectively.

\aa

The consistency of $(13)$ and  $(15)$ requires that
$$
{{\sum_S (2S + 1) H_\ell(N,S) \; 
{1\over 2} 
\left [
{N\over 2} ({N\over 2} + 1) - S(S + 1)
\right ]
}\over
{\sum_S (2S + 1) \; H_\ell (N ,S)}
} =
{{\ell + 1}\over {4\ell + 1}} \;
{{N(N - 1)} \over 2},
\eqno (23)
$$
from which we easily deduce
$$
< {\hat S}^2 > \; = {{3N}\over 4} \; 
\left (
1 - {{N - 1}\over {4\ell + 1}}
\right ),
\eqno (24)
$$
where $< {\hat S}^2 >$ is the average, over all the states of the 
configuration $n\ell^N$, 
of the square ${\hat S}^2$ of the total spin angular momentum. The
formula so-obtained for $< {\hat S}^2 >$ agrees with the one it is possible to
derive from
$$
< {\hat S}^2 > \; =
{{{\sum_S} (2S + 1) \; H_\ell(N,S)\; S(S + 1)}
\over
{
\pmatrix{
4\ell + 2 \cr
N \cr
}}}
\eqno (25)
$$
by using the explicit expression for 
$H_\ell (N,S)$ given in eqs.~(8) and $(10)$.

\aa
\aa

\noindent{\bf 4. Concluding remarks}

\sa

In this paper we have concentrated on electrostatic energy averages 
for $n \ell^N$ configuations in a 
new parametrization, viz., the ${\cal E}^\lambda (\ell, \ell')$ 
parametrization defined by (2) and (3). It is to be emphasized that 
the obtained averages (13) and (15) 
depend only on two parameters (${\cal S}$ and ${\cal D}$). This result and 
the fact, already noted in ref.~[6], that the term energies for the 
configuration $n \ell^2$ assume a very simple form in the 
${\cal E}^\lambda (\ell, \ell')$ parametrization, are two indications that 
a hidden symmetry is probably inherent to the 
${\cal E}^\lambda (\ell, \ell')$ parametrization. In this respect, it would 
be interesting to find a group theoretical interpretation for the 
${\cal E}^\lambda (\ell, \ell')$ parametrization.

\aa

In addition to the well-known interest, mentioned in refs.~[1] and [2], of 
energy averages, it is to be pointed out that compact formulas, like (13) 
and (15), for such averages constitute useful means for checking energy 
matrices. Furthermore, eq.~(15) suggests a strong version of Hund's 
rule, according to which the energy average 
$^{2S + 1}E_{av}(\ell , \ell)$ over all the states having a 
given total spin $S$ decreases (linearly) in $S(S+1)$ upon increasing $S$. 
This statement depends on the fact that the coefficient of $S(S+1)$ in (15) 
should be negative. It would be of interest to examine the effect that an 
independent optimization of the radial wave\-function for the 
average energy 
corresponding to each spin $S$ will have on the functional form 
of the dependence of this energy on $S(S+1)$ (see also ref.~[11]). 

\aa

The result (15) concerns the electrostatic interaction, 
in the ${\cal E}^\lambda (\ell, \ell')$ parametrization, within an 
$n\ell^N$ configuration in spherical symmetry ($SL$ coupling). This 
suggests several 
possible extensions of some of the results contained in the present 
paper. In particular, it would be appealing to extend the 
${\cal E}^\lambda (\ell, \ell')$ parametrization to other interactions 
(e.g., to the general interaction described by (1)) and to other 
configurations in arbitrary symmetry (e.g., to atomic and 
nuclear configurations 
with several open shells in spherical symmetry or to the molecular 
configuration $a_{2u}^{N_1} \, t_{1u}^{N_2} \, t_{2u}^{N_3}$ 
in cubical symmetry). 
Along this vein, it is to be mentioned that electrostatic energy 
averages have been derived, in the $F^{(k)}(\ell, \ell')$ parametrization, 
for $j^N$ configurations in spherical symmetry ($jj$ coupling) [12]. 

\aa

To derive (15) we used (10) which furnishes a new way for obtaining the 
number of states in $n \ell^N$ with a given spin $S$ in the case of 
spherical symmetry ($G = SO(3)$). It would be worthwhile to extend (10), 
and more geneally the generating function 
(4), to the case of an arbitrary point symmmetry group $G$. It would be also 
very useful to extend (4) to other systems (e.g., quark systems) than 
systems involving electrons by introducing in 
(4) additional degrees of freedom 
(e.g., isospin, flavor and color).

\aa

The energy averages (21) and (22) actually were obtained by inspection of 
tables giving term energies for $n\ell^N$ in the 
${\cal E}^\lambda (\ell, \ell')$ parametrization. As is often the case in 
spectroscopy, some regurality in energy formulas is the signature of 
special relations between coupling and/or recoupling coefficients. In this 
regard, we showed that the compact formulas (21) and (22) 
are a direct consequence of the sum rules 
for Clebsch-Gordan coefficients derived in 
the appendix.

\aa
\aa

\noindent {\bf Acknowledgement}

\sa

This work was elaborated during the visit of one of the authors (M.~K.)
to the Department of Chemistry of Technion (Israel Institute of Technology). 
This author would like to thank the Department of Chemistry for 
providing him with a visiting professor position and for the kind hospitality 
extended to him during his visit at Technion.

\aa
\aa

\noindent {\bf Appendix. Sum rules for $3-jm$ symbols}

\aa

The aim of this appendix is to report two apparently new sum rules for $SU(2)
\supset U(1)$ coupling coefficients. Equations (21) and $(22)$ of the main
body of this paper are simple consequences of these sum rules.

\aa

Let us consider the decomposition
$$
(\ell) \otimes (\ell) = \{ (\ell) \otimes (\ell) \}_+ 
\bigoplus \, [(\ell) \otimes (\ell)]_-,
$$
$$
\{ (\ell) \otimes (\ell) \}_+ = 
\mathop{\bigoplus}\limits_{L_e = 0 (2) (2 \ell)} 
(L_e), \qquad [(\ell) \otimes (\ell)]_- =
\mathop{\bigoplus}\limits_{L_o = 1(2) (2\ell - 1)} (L_o),
\eqno (26)
$$
into a symmetrized part $\{ \ \ \}_+$ and an 
  anti-symmetrized part $ [ \ \  ]_-$, of the direct product 
$(\ell) \otimes (\ell)$ of two irreducible
representation classes $(\ell)$ of $SU(2)$. Such a decomposition can be
transcribed in terms of basis vectors, acting on two spaces (1 and 2) of
constant angular momentum $\ell$, as 
$$
|\ell m)_1 \otimes |\ell m')_2 + \pi \; |\ell m')_1 \otimes |\ell m)_2 
= 2 \; \sum_{L_\pi M} \; |\ell \ell \; L_\pi M) \; (\ell \ell
mm' | L_\pi M),
\eqno (27)
$$
where
$$
L_\pi \equiv L_e = 0 (2) (2\ell) \quad {\rm for} \quad \pi = +1 
                              \quad {\rm and} \quad 
L_\pi \equiv L_o = 1 (2) (2\ell - 1) 
                              \quad {\rm for} \quad \pi = -1.
\eqno (28)
$$
By taking the scalar product of $(27)$ with $|\ell \mu)_1 
\otimes |\ell \mu')_2$, we get the sum rules
$$
\delta (m,\mu) \delta (m',\mu') +
\pi \; \delta(m',\mu)\;
\delta(m,\mu') =
2 \; \sum_{L_\pi M} \;
(\ell \ell \; \mu \mu' | L_\pi M)\;
(\ell \ell \; mm' | L_\pi M)
\eqno (29)
$$
or, equivalently, in terms of $3-jm$ symbols
$$
\sum_{L_\pi M} 
(2L_\pi + 1) 
\pmatrix{
L_\pi &\ell &\ell\cr
-M & m &m'\cr
}
\pmatrix{
L_\pi &\ell &\ell\cr
-M &\mu &\mu'\cr
}
= {1\over 2} 
\left [
\delta (m,\mu)
\delta (m',\mu') + \pi \delta (m', \mu)
\delta (m, \mu')
\right ].
\eqno (30)
$$
The particular case $(m = \mu = 0$, $m' = \mu' = \lambda)$ leads to (21) and 
$(22)$ for $\pi = + 1$ and $\pi = - 1$, respectively.

\vfill\eject

\noindent{\bf References}

\sa

\parindent = 0.8 true cm
  \baselineskip = 0.82 true cm

\item{[1]} J.~C. Slater, Phys. Rev. 34 (1929) 1293 ; Quantum theory of atomic 
structure (McGraw-Hill, New York, 1960).  

\sc

\item{[2]} E.~U. Condon and G.~H. Shortley, The theory of atomic 
spectra (Cambridge 
University Press, Cambridge, 1935) ; 
G.~H. Shortley, Phys. Rev. 50 (1936) 1072 ;  
G.~H. Shortley and B. Fried, Phys. Rev. 
54 (1938) 739. For a recent comprehensive review of 
atomic structure, see : E.~U. 
    Condon and H. Odaba\c s\i,   
Atomic structure (Cambridge University Press, Cambridge, 1980). 

\sc

\item{[3]} G. Racah, Phys. Rev. 61 (1942) 186 ; 
                      62 (1942) 438 ; 
                      63 (1943) 367 ; 
                      76 (1949) 1352. 

\sc

\item{[4]} R. Glass, Comput. Phys. Commun. 16 (1978) 11. 

\sc

\item{[5]} M. Kibler and G. Grenet, Phys. Rev. B 23 (1981) 967 ; Int. 
J. Quantum Chem. 29 (1986) 485.

\sc

\item{[6]} M.~R. Kibler, Int. J. Quantum Chem. 9 (1975) 421. 

\sc

\item{[7]} C.~K. J\o rgensen, R. Pappalardo and H.-H. Schmidtke, J. Chem. Phys. 
39 (1963) 1422 ; 
C.~E. Sch\"affer and C.~K. J\o rgensen, Mol. Phys. 9 (1965) 401.

\sc

\item{[8]} M. Kibler, J. Chem. Phys. 61 (1974) 3859.

\sc

\item{[9]} J. Katriel and A. Novoselsky, J. Phys. A : Math. Gen. 22 
(1989) 1245.

\sc

\item{[10]} C.~W. Nielson and G.~F. Koster, Spectroscopic coefficients for the 
$p^n$, $d^n$, and $f^n$ configurations (M.~I.~T. Press, Cambridge, Mass., 
1963).

\sc

\item{[11]} J. Katriel and R. Pauncz, Adv. Quantum Chem. 10 (1977) 143.

\sc

\item{[12]} A. de-Shalit and I. Talmi, Nuclear shell theory (Academic, 
New York, 1963).

\bye